\documentclass[english,showpacs,twocolumn]{article}
\usepackage[utf8]{inputenc}
\usepackage{graphicx}
\usepackage{changepage}
\usepackage{amsmath,amssymb}
\usepackage{authblk}

\newcommand\dblquote[1]{\textquotedblleft #1\textquotedblright}

\usepackage{soul}
\usepackage{comment}
\usepackage[T1]{fontenc}
\usepackage{lmodern}

\begin{document}

\title{Statistical Physics of Balance Theory}
\author[1,2]{Andres M. Belaza}
\author[2]{Kevin Hoefman}
\author[1]{Jan Ryckebusch}
\author[2,3,4]{Aaron Bramson}
\author[1,2]{Milan van den Heuvel}
\author[2,5]{Koen Schoors}
\affil[1]{Department of Physics and Astronomy, Ghent University, Ghent, Belgium}
\affil[2]{Department of General Economics, Ghent University, Ghent, Belgium}
\affil[3]{Laboratory for Symbolic Cognitive Development, RIKEN Brain Science Institute, Wakoshi, Saitama, Japan}
\affil[4]{Department of Software and Information Systems, University of North Carolina Charlotte, Charlotte, North Carolina, USA}
\affil[5]{Higher School of Economics, National Research University, Moscow, Russia}
\date{}

\twocolumn[
  \begin{@twocolumnfalse}
    \maketitle
 
\begin{abstract}

Triadic relationships are accepted to play a key role in the dynamics of social and political networks.  Building on insights gleaned from balance theory in social network studies and from Boltzmann-Gibbs statistical physics, we propose a model to quantitatively capture the dynamics of the four types of triadic relationships in a network. Central to our model are the triads' incidence rates and the idea that those can be modeled by assigning a specific triadic energy to each type of triadic relation. We emphasize the role of the degeneracy of the different triads and how it impacts the degree of frustration in the political network. In order to account for a persistent form of disorder in the formation of the triadic relationships, we introduce the systemic variable temperature. In order to learn about the dynamics and motives, we propose a generic Hamiltonian with three terms to model the triadic energies. One term is connected with a three-body interaction that captures balance theory. The other terms take into account the impact of heterogeneity and of negative edges in the triads. The validity of our model is tested on four datasets including the time series of triadic  relationships for the standings between two classes of alliances in a massively multiplayer online game (MMOG).  We also analyze real-world data for the relationships between the ``agents'' involved in the Syrian civil war, and in the relations between countries during the Cold War era.  We find emerging properties in the triadic relationships in a political network, for example reflecting itself in a  persistent hierarchy between the four triadic energies, and in the consistency of the extracted parameters from comparing the model Hamiltonian to the data.  

\end{abstract}
\vspace{1cm} 
\end{@twocolumnfalse}
]

\section{Introduction}

Signed social networks are those with both positive and negative edge weights used to capture the valence as well as the strength of dyadic relationships, such as friendship/ally and animosity/enemy. By far the dominant method to analyze such networks is balance theory (also called ``structural balance theory'').  Originally proposed by Heider \cite{heiderattitudes} as an explanation for attitude change, and formally generalized for graphs by Cartwright and Harary \cite{cartwright1956sbg}, balance theory has been refined and applied to a variety of social, economic, ecologic and political scenarios \cite{Hart1974,Hummon2003,Leskovec2010,Doreian2015,Lerner2016,Marvel2009,Antal2005,Saiz2017}. Balance theory provides a means to capture a system of such relationships and measure the degree to which it is balanced/stable or frustrated/unstable.  Here we provide a further expansion of balance theory utilizing methods from Boltzmann-Gibbs statistical physics to show that after assigning energy levels to unique configurations of triads in signed graphs, we can extract the characteristic strength parameters of their interactions. 

The original formulation was concerned only with whether graphs were balanced or not, where ``balanced'' meant that all cycles among all nodes contained only an even number of negative edges \cite{cartwright1956sbg}.  Later work \cite{Abell1968} argued that only triads of nodes (i.e., connected triples, 3-cycles) were relevant for most social science applications of balance theory, and also showed that the proportion of unbalanced $n$-cycles and 3-cycles increase monotonically with each other. Thus the triadic version has become dominant (although see\mbox{\cite{Facchetti2011}} and\mbox{\cite{Estrada2014}} for alternate measures of balance). Balanced configurations are still those with an even number of negative edges; specifically we capture the ideas of ``The enemy of my enemy is my friend'' $[+--]$ and ``The friend of my friend is also my friend'' $[+++]$ as balanced/stable situations.  The two other types of signed triadic configurations ($[++-]$ and $[---]$) are considered unstable and give rise to frustration in the network. Because this was developed as a theory of attitude change, the frustrated triads are considered to be posed to change to increase systemic balance.

As a further refinement, social scientists since \cite{Davis1967} have observed that the two types of balanced triads are not equally balanced, and the two types of frustration are not equally frustrated.  If we consider the classic version above the ``strict rule'' for balance, the ``loose rule'' for balance rates $[+++]$ as more strongly balanced than $[+--]$, and $[++-]$ is more strongly frustrated than $[---]$.  This breakdown reflects the observation that while triple negative triads are not stable, they do not actually inject much frustration into the system.  Likewise, although $[+--]$ is balanced, a system containing entirely triple positive triads is more stable.  


\begin{figure}[!h]
\centering
\includegraphics{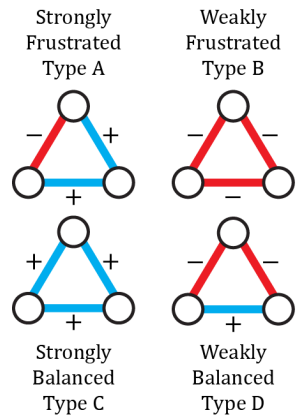}
\caption{{\bf Different types of triadic relations and their classification according to the strict and loose rule.} In both classifications, C and D are balanced. According to the strict rule, the triads A and B are equally unbalanced. According to the loose rule, A is more unbalanced than B.}
\label{fig:type_triads}
\end{figure}


Heider's original hypothesis was that a network of signed relations would tend to evolve towards a more balanced situation, eventually having solely triple positive balanced triadic relationships (often referred to as ``utopia'').  Several simulation models explore the formal conditions for this outcome~\cite{cartwright1956sbg,Antal2006,ABELL2009,Gawroski2005,Traag2013}, but empirical studies reveal significant deviations.  The validity of SBT has been tested on a variety of social and political relationships, such as political relations between countries \cite{Hart1974,Doreian2015,Lerner2016}.
For example, in the time span between 1946 and 1999 the fraction of unbalanced triads in the political network studied by \cite{Doreian2015} fluctuates over time and is consistently in the 5-15\% range. Furthermore, there was no decreasing trend in the fraction of unbalanced triads. Although the rationale behind Heider's theory is compelling, and there may be intervening and exogenous factors (such as those explored by \cite{Lerner2016}) that explain the discrepancy, the power of balance theory to explain political dynamics remains an open question.

One recurring problem in the study of social and political systems is that the data are incomplete and/or subject to privacy-related restrictions.  Considering this, the virtual worlds of massively multiplayer online games (MMOGs) are fascinating social laboratories that can serve as sources for high-quality data in a controlled environment over extended periods of time.  For example, the authors of \cite{Szell2010} infer the signed social network in the online virtual world PARDUS using information available such as the formation of alliances, trade, attacks, the exchange of private messages, etc.  They found different structural properties in the positive and negative networks with respect to clustering coefficients, reciprocity, and degree distributions. They also found that the two types of unbalanced triads have a different incidence rate; specifically the $[---]$ (weakly frustrated) triads are more frequent than the $[++-]$ (strongly frustrated) ones. The two types of balanced triads also have a markedly different incidence rate; specifically the $[+++]$ (strongly balanced) triads have a higher frequency than the $[+--]$ (weakly balanced) ones. Aside from implying that cooperation between the players is an important incentive in PARDUS, this provides support for using the loose version of balance theory.

For our main analysis we use data on the ally/enemy relationships between two different classes of alliances from the virtual world of EVE Online (another MMOG). These data are aggregated daily and cover a time period of over a year. Because these datasets come from a virtual world, complete and accurate records of all alliance standings across time are available. We also analyze two datasets from the real world: (1) the relations among fourteen political participants in the military intervention against ISIS (DAESH), and (2) the relations between states during the Cold War era.

Pioneering work in applying physics-based models inspired by the Ising model to political networks can be found in \cite{Galam1996} and \cite{Vinogradova2014}. In these works, the spin variable is used to  assign a political agent to a peculiar global alliance. Thereby the spins are also  subject to history-based interactions. We develop an alternate approach based on balance theory.

Our method applies principles of Boltzmann-Gibbs statistical physics to balance theory.
First, we test the accuracy of the traditional strict and loose rules of balance theory. Next, we argue that the incidence rate (or, occupation probabilities) for the four types of triadic relationships are good measures to gain a better understanding of balance theory.   We associate a specific energy value to each type of triad (``triadic energies'') and introduce the concept of \textit{temperature} as a measure of the total and persistent systemic frustration. The information entropy corresponding to the occupation probabilities of triads is introduced. In order to unravel the dynamics behind structural balance, we introduce a Hamiltonian with three parameters that can be extracted from model-data comparisons. We find a high degree of consistency among the model parameters extracted from the four datasets analyzed in this article, thus implying that our statistical physics approach to balance theory may be broadly useful.

\section{Materials and Methods}
\label{sec_statphysofSBT}
In this section, we capture social balance in a statistical physics framework. Previous formal approaches to analyzing systemic balance and dynamics often focus on changing the link values or structure to understand tendencies towards or away from balance. Antal et al.~\cite{Antal2006} proposed a model in which randomly selected links change with the aim of balancing unbalanced triads. In this model, dubbed ``local triad dynamics'', they found that a finite network relaxes into an equilibrium state with  balanced triads.  Abell and Ludwig explore the tradeoff between increasing the number of positive links and the nodes' tolerance to imbalance. Variations in these parameters provided evidence for three behavioral phases, including features resembling self-organized criticality~\cite{ABELL2009}. Gawronski et al.~use continuous link values to reformulate balance theory in terms of dynamical equations. They show that, given certain constraints, in a fully connected network the system converges to Heider's balanced state (or ``utopia'') after a finite number of interactions~\cite{Gawroski2005}.
  
A formulation of balance theory in terms of energy levels using only the strong rule was proposed by~\cite{Marvel2009}, but its focus was on explaining why the systems do not necessarily evolve to lower frustration. They did this by investigating the landscape of possible networks and found so-called ``jammed states'' that impede the total relaxation of the system.  Although balance theory includes a tendency toward reduced frustration, we are also interested in various systems' tolerance levels of systemic frustration.

We propose a framework allowing one to determine the proportions of triad types a given system adopts. To do this we assign a characteristic energy to each of the four triad types in Fig.~\ref{fig:type_triads}, which produces a ranking of energy levels.  Those so-called ``triadic energies'' determine how many of each type of triad persists in the system.  
In this framework, low-energy triads will be more common than what would be expected based on their rate of occurrence in a random network. The opposite is true for high-energy triads. By comparing various systems one can infer whether some robust properties  emerge with respect to the frequencies of each type of triad. 

\subsection{Energies and Entropies of Political Networks.}

Political networks are dynamic, so the proportion of each type of triad and the density of the triads will fluctuate over time. This includes triads among particular triplets of nodes that change type, an effect that can be modeled by considering that it changes energy level. In our applications the system never resides in its energetically most favored state (``utopia'' or the``ground state''), as is commonplace in physical system with a non-vanishing temperature. The social network dynamics of our systems can be attributed to several sources (conflict, political bargaining, access to resources, trading partners, etc.), but we will not discriminate between these different drivers of change. We model their overall (averaged) effect through the concept of ``temperature''. This can be seen through fluctuations in quantities such as the average number of triads of a particular kind around the mean.

Now, we are going to apply the methodology of classical statistical physics to the study of social networks, focusing on the triadic relations as the level where the preferences of the different nodes play the main role. First, we need to define the variables that characterize the system. A social network of $N$ nodes is composed of $\frac{1}{2}N(N-1)$ possible symmetric edges $s_{ij}$ between nodes $i$ and $j$, each of which must have one of three values: $s_{ij} \in \{+, - , \times \}$ (positive, negative or nonexistent).  Thus a microstate $\mu$ of such a network is uniquely specified by the complete set of $s_{ij}$ values (together with any relevant node attributes). 

\begin{figure}[tb]
\centering
\includegraphics{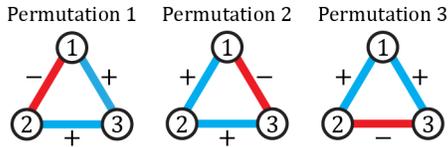}
\caption{{\bf Geometrical degeneracy of triads.}
The $[++-]$ triad (Type A) has 3 different micro-states defined by the permutations of edge valences shown here.  These micro-states constitute the geometrical degeneracy of the triads.
}
\label{fig:geometrical_degeneracy}
\end{figure}

Our unit of analysis here is triads, specifically the four types of triads presented in Fig.~\ref{fig:type_triads}. We define a triad micro-state as the set of $\frac{1}{6}N(N-1)(N-2)$ triads $T_{ijk}$ between nodes $i$, $j$ and $k$, each of which must have one of five states: $T_{ijk} \in \{A,B,C,D,\times \}$ (The four kinds of Fig.~\ref{fig:type_triads} or nonexistent). The map from network micro-states $\mu$ to triad micro-states $\eta$ is not bijective: all micro-states $\mu$ correspond to a unique triad micro-state $\eta$, but not vice-versa. 
The states of two triads that share a link are not actually completely independent, but taking the potential correlations into account is notoriously challenging. For example, although there are no shared links between the $[+++]$ and $[---]$ triads, a negative edge may be part of triads of type A, B, or D in various permutations. For sparse networks the number of triads sharing links is low and so, to a first-order approximation, the triads can be considered as independent.
From the perspective of the triads, edge permutations can be seen as the geometrical degeneracy of the triads, and the number of equivalent edge microstates differs for the various types of triads.  Type B and C triads have three identical edges, so there is only one edge combination that can generate those.  Triad types A and D can be generated in three different ways depending on which edge is the odd one out (see Fig.~\ref{fig:geometrical_degeneracy}). 


We must first estimate the probabilities of each type of triad: $p_{A}$, $p_{B}$, $p_{C}$, $p_{D}$. Because these types constitute a partition of triadic micro-states, we can divide the number of each type $m_i$ by the total number of triadic relations in the network $M$ to produce proportions.  We make the simplifying assumption that the probabilities can be estimated accurately by these observed proportions: 
\begin{equation}
p_{i\in\left\{A,B,C,D \right\}} = \frac {m_i}{M} \;.  
\label{eq:occupprobability}
\end{equation}
Unlike the case of the network in which each unique configuration of the edges counted as a distinct micro-state, for the triads we are considering a micro-state as defined by the \emph{proportion} of each type of triad. This means that the micro-states $\mu$ are uniquely determined by the variables $\left\{p_A,p_B,p_C,p_D \right\}$. In this procedure, we aggregate both the geometrical degeneracy within each type as well as which particular triads are in which particular configuration.  We do this because we are interested in the total average energy and entropy of the system, and such macro-variables are determined by a weighted sum over all possible micro-states. 

We assign an energy $E_i$ to each type $i$ of triad. The corresponding degeneracy $g(E_i)$ is defined as the total number of different ways of creating a triad of type $i$. Apart from the geometrical degeneracy (see Fig.~\ref{fig:geometrical_degeneracy}), there can be other sources contributing to $g(E_i)$.  The triadic entropy $S_T$ generated by the probabilities $\bigl\{ p_{A}, p_{B} ,p_{C},p_{D} \bigr\}$ of the energy levels of each type of triad can be calculated as
\begin{equation}
S_T =- K\sum_{i\in\left\{A,B,C,D \right\}} p_i \left( \ln{p_i} - \ln{g(E_i)} \right) \; \; ,
\label{eq:entropy_occupation}
\end{equation}
where $K$ is a constant that connects the units of entropy and energy. This constant is the  Boltzmann's constant ($k_B$) in statistical physics.

We now derive an expression for the probabilities $p_{i\in\left\{A,B,C,D \right\}}$. If we assume that each type of triad can be connected to a specific energy, the principle of maximum entropy applied to the $S_T$ under constraints of normalization of the probabilities and a finite average energy leads to the Boltzmann distribution \cite{Presse2013}. Thereby, each energy level $i \in \left\{ A,B,C,D \right\}$ has a probability of occupation $p_i$ given by
\begin{equation}
p_{i \in \left\{ A,B,C,D \right\}}=\frac{g(E_i) \;  \exp(-\frac{E_i}{KT})}{Z} \; ,
\label{boltzmann}
\end{equation}
whereby the partition function 
\begin{equation}
{Z}=\sum_{i \in \left\{ A,B,C,D \right\} } g(E_i) \;  \exp(-\frac{E_i}{KT}) \; ,
\label{eq:partifunc}
\end{equation}
acts as a normalization constant to ensure the total probability $\sum_{i \in \left\{ A,B,C,D \right\} } p_i = 1$. Further, $(KT)$ has the units of energy and is the Lagrange multiplier associated with the constraint that the system has an average total energy. In the remainder of this work, we also use the notation $\beta \equiv \frac {1}{K T}$.  From this formulation we create a model in which, for a ``high'' temperature (defined as $ KT >> \left|E_i - E_j \right| \forall i,j$), the triads will be stochastically occupied according to their degeneracy $g(E_i)$, because $\exp(-\frac{E_i}{KT})\approx \exp(-\frac{E_j}{KT}) \forall i,j$. For ``small'' temperatures (defined as $ KT << \left|E_i - E_j \right| \forall i,j$) all triads will be in the lowest possible energy state $E_g$, because $\exp(-\frac{E_g}{KT})>>\exp(-\frac{E_j}{KT})$. In the parlance of balance theory, this specific small-temperature situation corresponds with utopia, a network of  exclusively $[+++]$ triads.

The expression of Eq.~\ref{boltzmann}  connects the occupation probability of a certain triadic state to the ratio of its energy and the temperature (so it is invariant under translations of the energy scale). Our methodology determines the triadic energies from the occupation probabilities; so although it is fundamentally impossible to determine the temperature, without any loss of generality we can set KT as the unit of energy.

We study the time evolution of social networks, which may be considered out-of-equilibrium systems. However, because our data are collected at time intervals that are short relative to the social dynamics, we interpret them as quasi-static. Consider that political networks that typically evolve over the time scales of years may be interpreted as quasi-static if the data are recorded on a daily basis. Thus Eq.~\ref{boltzmann} provides the triadic occupation probabilities for such quasi-static out-of-equilibrium scenario. 

In a political network with agents not displaying preferences, the relative occupation probabilities are determined solely by the degeneracies. The rate of incidence of the different types of triads is then a reflection of the degeneracy. When the agents display preferences --which is reflected in a certain hierarchy of energy levels --, the  degeneracy still plays a major role. A level with a relatively high energy, which means that it is not the preferred state of the agents, could still have a large incidence if its degeneracy is large enough. Random networks can be used as a reference for the preferences among the different energy levels.  In a study with random networks, Szell et al.~\cite{Szell2010} report relative occupation probabilities in their random network for the triads B:A:D:C close to 1:3:3:1 which reflects their geometrical degeneracy. Using triadic relationships produced by random signed networks as a control, we can detect which triads are conceived of as more unbalanced (the ones underrepresented in the data relative to random networks) or more balanced (overrepresented relative to random networks). 

Given data for the occupation probabilities $p_i$ of the different types of triads, the relative energies $\left( E_i - E_j \right)$ can be extracted from
\begin{equation}
\beta E_i - \beta E_j  = \left[ -\ln \frac{p_i}{g(E_i)} - \ln Z \right] - \left[ 
 - \ln \frac{p_j}{g(E_j)} - \ln Z \right] \; .
\label{eq:Energy_from_ocupation_probability}
\end{equation}
Throughout this article we systematically refer to the quantity ``$-\ln \frac{p_i}{g(E_i)}  $'' as the ``extracted triadic energy'' $\beta E_i + \ln Z$ for triad type $i$. 
Using the $\beta$ as a reference of the energy scale and the inferred degeneracy of each type of triad, we have developed a methodology that allows us to solve for the relative energies between the triad types based on their frequency of occurrence in data of a political network. As we will see shortly, the energies can be connected with the underlying dynamics. 

\subsection{Hamiltonian for the Dynamics of Triadic Relationships}

The use of the combination of degeneracy and energy for each specific triad will allow us to embark on quantitative studies of the underlying dynamics of political networks. Up to this point we have connected the rate of incidence of a specific triad $i\in \left\{ A,B,C,D \right\}$ to its energy $E_i$ (different for each state, and displaying the preferences) and a temperature $KT$ (which is a systemic variable). 
The authors of \cite{Marvel2009} showed that strict-rule balance theory can be mapped into a genuine three-body interaction among the edges because the product of the three edge valences discriminates between balanced and unbalanced states.The three-body interaction solely differentiates between balanced and unbalanced triads. We wish to further differentiate between strongly and weakly balanced/frustrated triads (Fig.\ref{fig:type_triads}).  To this end, we add a one-body and a two-body part to the Hamiltonian. In order to gain a more detailed insight into the underlying dynamics, we propose to add a two-body and a one-body interaction between the edges. The effects of the three-body, two-body, one-body terms are regulated by strength parameters $\alpha$, $\gamma$, and $\omega$ respectively. The three contributions to the interactions between the nodes are designed such that the values of those parameters $\left( \alpha, \gamma, \omega \right)$ provides detailed information about the node's incentives with regard to establishing relationships.  

Let $s_{ij} \in \{ -1, +1 \}$ encode the relationship (enemy or friend) between nodes $i$ and $j$.  Then the proposed Hamiltonian acting in the space of a given number $M$ of complete triads is

\begin{align}
\begin{split}
\mathcal{H}= \sum_{i<j<k} & [ \underbrace{-\alpha s_{ij} s_{jk} s_{ki}}_{\text{three-body}} 
\\ & \underbrace{- \gamma (s_{ij}s_{jk}+s_{ij}s_{ki}+s_{jk}s_{ki})}_\text{two-body} 
\\ & \underbrace{+ \omega (s_{ij} + s_{jk} + s_{ki})}_{\text{one-body}}
] \; .
\label{eq:hamiltonian}
\end{split}
\end{align}

\begin{figure*}[tb]
\begin{center}

\includegraphics{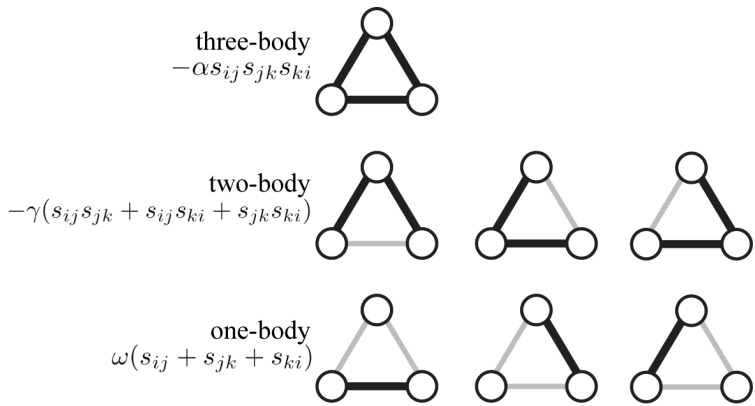}
\caption{\textbf{Schematic diagram illustrating the three terms in the proposed Hamiltonian of Eq~\ref{eq:hamiltonian}.}
} 
\label{fig:Hamiltonian_and_substructures}
\end{center}
\end{figure*}

Here, the sum $\sum_{i<j<k}$ extends over all unique 3-node cycles $\left\{1,2, \ldots, M \right\}$ in the network.  The indices $i,j,k$  run over nodes that are part of complete triads. In this way the Hamiltonian for each triad type is derived directly from the valences of the edges that make them up and shown in Table~\ref{table:Levels}. Fig.~\ref{fig:Hamiltonian_and_substructures} shows a schematic diagram of the three terms in the proposed Hamiltonian. 

\begin{table*}[tb]
\centering
\begin{flushleft}
\caption{\textbf{Energies and degeneracies associated with the four types of triadic relationships  considered in this work.} \label{table:Levels} }
\begin{tabular}{ l l l l} 			
&
            & \multicolumn{1}{c}{Associated energy}
            & \multicolumn{1}{c}{Geometrical}  \\
            \multicolumn{1}{c}{Type of triad}
            & \multicolumn{1}{c}{Symbol}   
            & \multicolumn{1}{c}{from Hamiltonian (\ref{eq:hamiltonian})}
            & \multicolumn{1}{c}{degeneracy} \\ \hline 
Highly frustrated & A: $[++-]$ &   $H_A = \alpha + \gamma +\omega$ & $g_G (E_A)=$3 \\ 
Lowly frustrated & B: $[---]$ & $H_B = \alpha - 3\gamma -3\omega$ & $g_G (E_B)=$1 \\
Lowly balanced & D: $[+--]$ & $H_D = -\alpha +\gamma -\omega $ & $g_G (E_D)=$3 \\  
Highly balanced & C: $[+++]$ & $ H_C = -\alpha -3\gamma +3\omega$ & $g_G (E_C)=$1 \\ 
\hline
\end{tabular}
\\The corresponding Hamiltonian in the space of triadic relationships is given by the Eq~(\ref{eq:hamiltonian}). The geometrical degeneracy is explained in Fig~\ref{fig:geometrical_degeneracy}.
\end{flushleft}

\end{table*}
The two-body term in the Hamiltonian with strength $\gamma$ can be interpreted as a force term that attempts to ``homogenize'' the relations in the triad. It introduces a fine-splitting of magnitude $4 \gamma$ in the energy spectrum depending on whether the balanced or unbalanced triad is symmetric or not (Table~\ref{table:Levels}). The parameter $\gamma$ makes clustering energetically more favorable, even if it concerns a triad with solely enemies.  The energy of the balanced and unbalanced triads are lowered by an amount $3\gamma$ for the symmetric $[+++]$ and $[---]$ triads and increased by an amount $\gamma$ for the asymmetric triads $[+--]$ and $[++-]$.  The strength $\omega$ of the one-body term in Eq.~\ref{eq:hamiltonian} encodes the ``reward'' that corresponds with the creation of ``$-$'' links. 

The values of the strength parameters can be determined by analyzing data of triadic relations in empirical networks, and doing so tells us about the underlying dynamics of such systems. To see how this works, we now demonstrate this statistical approach to structural balance theory with a few example applications.

\subsection{Virtual World Data: Political Network of Alliances in EVE Online}
\subsubsection{Data}

The extended dataset that we first analyze is extracted from recorded and time-stamped data from EVE Online, which is an MMOG developed by CCP Games~\cite{EveOnline}. In this virtual world more than 500,000 players trade, collaborate and fight in a futuristic galaxy. The players get together in social structures called alliances with sizes between one and about 25,000 players. The alliances can conquer territory, where they can impose their own taxes, exploit mineral resources, and so on. The relations between the alliances represent an important social aspect of the game including the fact that the leader of the alliance has four choices with regard to setting the relationship to the other alliances in the game.  Indeed, the relationship to any other alliance can be set as friendly, hostile,  neutral or undetermined. This is important because it will change how the players of one alliance deal will the players in the other alliances, facilitating the process of discriminating between friends, enemies, and others.  

Although one might expect a symmetric relationship for the standings between the alliances 1 and 2 $\left( s_{1 \rightarrow 2} , s_{2 \rightarrow 1} \right)$, the data reveal a slight degree of asymmetry in the directed network. The degree of reciprocity (having the same link valence in both directions) is in the range 0.9-0.98 across our time series, so the assumption of symmetry is still largely justified. Because triadic balance theory works best on undirected edges, we transform the directed edge standings data into symmetric relationships via the conversion rules summarized in  Table~\ref{table:PTM}. These conversion rules are inspired by the rules of the game so that the status of the relationship between the alliances really reflects the dynamics of EVE Online.

\begin{table}[t]
\caption{\textbf{Conversion table for the network of alliance relations in EVE.} \label{table:PTM}}
\begin{flushleft}
\begin{tabular}{ l l | c | c }
\multicolumn{1}{c}{$s_{1 \rightarrow 2}$} & \multicolumn{1}{c}{$s_{2 \rightarrow 1}$} & $ s_{12} $ & Multiplicity \\ 
			Not Set & Not Set &  $\times$ & 1\\  
			Not Set & Friend & $+$  & 2  \\  
			Not Set & Neutral &  $\times$ & 2 \\  
			Not Set & Enemy & $-$  & 2 \\  
			Friend & Friend & $+$ & 1 \\   
            Friend & Neutral & $+$ & 2  \\  
			Friend & Enemy & $-$  & 2 \\   
			Neutral & Neutral &  $\times$ & 1   \\   
			Neutral & Enemy & $-$ & 2 \\ 
			Enemy & Enemy & $-$ & 1 \\  
            \end{tabular} 
\\ The rules used to transform the directed network of alliance relations $\left(s_{1 \rightarrow 2}, s_{2 \rightarrow 1} \right)$ in EVE Online into an undirected network of alliance relations $s_{12}$. The multiplicity is 1 for situations whereby $s_{1 \rightarrow 2} = s_{2 \rightarrow 1}$ and is 2 otherwise.
\end{flushleft}

\end{table}

The dataset under analysis consists of a time series of the relations between the alliances between February 15, 2015 and April 16, 2016. We consider  \dblquote{politically active} alliances, defined by the criterion that they set  their standings with at least one other alliance. We study two distinct classes of alliances. First, the class of alliances with more than 200 members (so called \dblquote{+200} alliances), and second, the class of  alliances that hold  sovereignty over at least one solar system (\dblquote{SOV} alliances for short).  These two classes of alliances are key to the political dynamics of the game. 
We proceed with providing information about the network structure of the two classes of alliances.

\subsubsection{Network Properties of the Alliances in EVE Online}

\begin{figure*}[tb]
\begin{flushleft}
\includegraphics{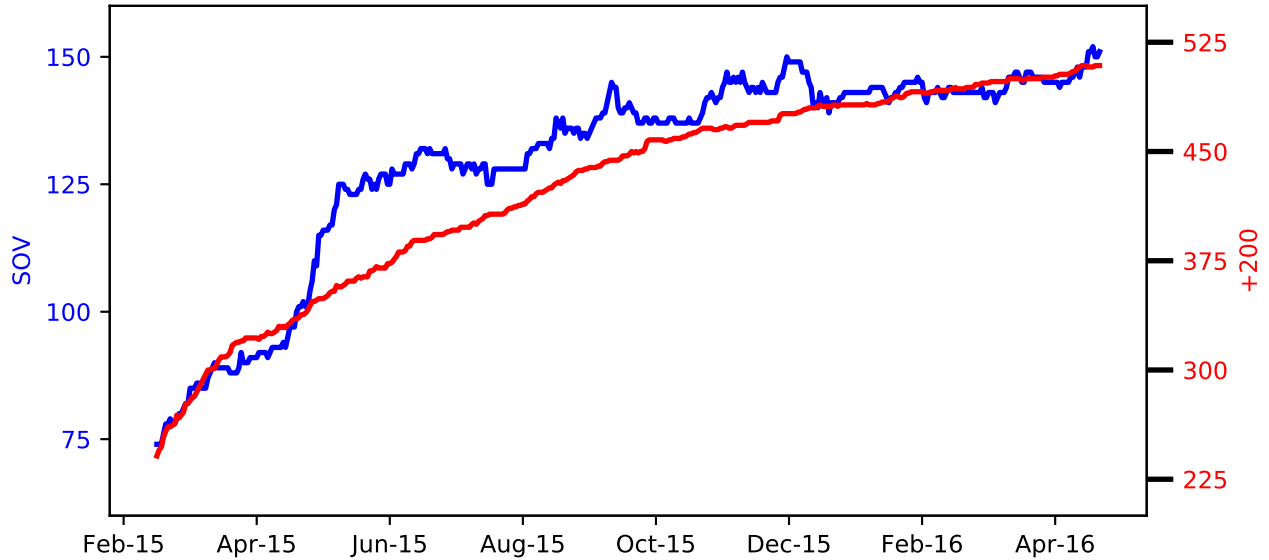}
\end{flushleft}
\caption{\textbf{Evolution of the number of nodes in the network of alliance relations.}
Daily evolution of the number of alliances in EVE Online from February 2015 through April 2016. We discriminate between alliances with sovereignty (SOV) and alliances with more than 200 members (+200)}
\label{fig:nodes_evolution}
\end{figure*}

Political networks are not static.
In EVE Online, new alliances come into the system and others disappear while the status of the relations between alliances are volatile as a result of all ongoing activities. In an effort to provide some feeling about the size of the data, we display in Fig.~\ref{fig:nodes_evolution} the time evolution of the number of alliances in the studied period. The number of SOV (+200) alliances grows from 70 to 150 (250 to 500) and is subject to temporal fluctuations on top of the growing trend. 

\begin{figure*}[htb]
\begin{flushleft}
\includegraphics{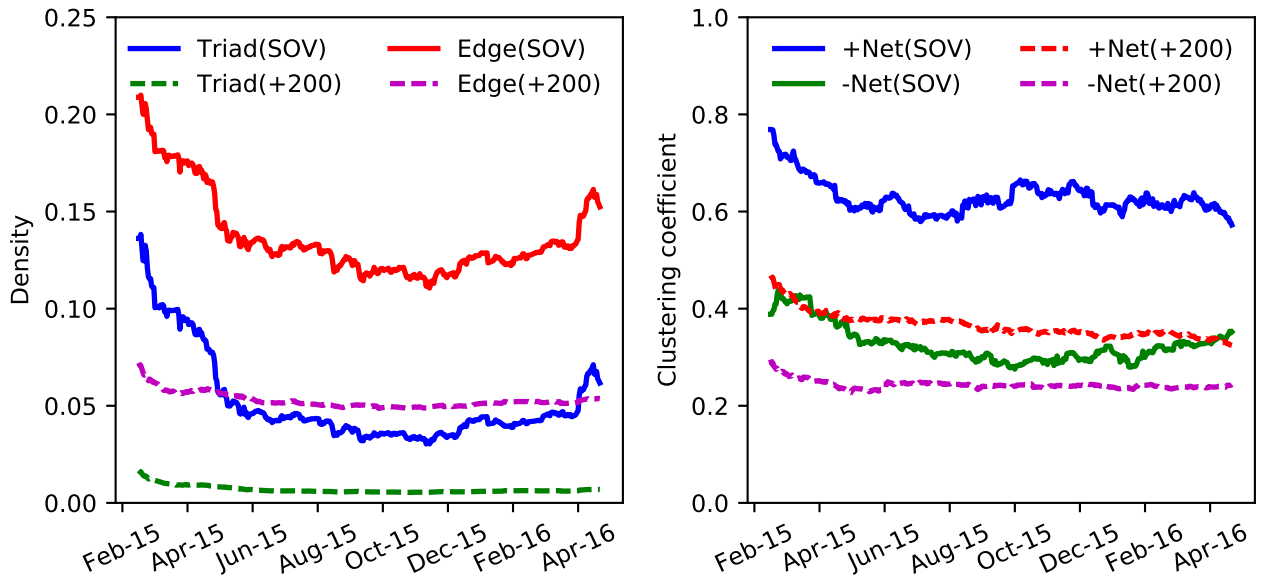}
\end{flushleft} 
\caption{\textbf{Properties of the network of EVE's alliances.}
{\bf Left:} Daily values of the density of complete triads and the density of edges for two classes of alliances in EVE Online. {\bf Right:} Daily values of the clustering coefficients of the network of positive (\dblquote{$+$Net}) and negative relationships  (\dblquote{$-$Net}) for the SOV and +200 alliances in EVE Online.} 
\label{fig:Network_Prop}
\end{figure*}

We now study the time evolution of some relevant network metrics of the SOV and +200 alliances. We focus on the evolution of the density of the triads, of the density of the edges and of the clustering coefficients. The results are summarized in  Fig.~\ref{fig:Network_Prop}. 
For the density of the triads and the edges there is a markedly different behavior of the SOV and the +200 alliances. The network of SOV alliances is marked by a higher density of edges and triads. We also observe strong temporal fluctuations in the densities of edges and triads for the SOV alliances. The density of complete triads fluctuates between 4\% and 14\%.  The big alliances have a smaller and rather stable density  of triads (order 1\%) and edges (5-6\%). Our analysis based on uncorrelated triads is a reasonable approximation for networks with such an observed low triad density.

For the clustering coefficients, we discriminate between the subnetworks of the hostile (\dblquote{$+$}) and friendly (\dblquote{$-$}) relationships.  Clearly, the SOV and +200 alliances display a markedly different behavior in their clustering coefficients. For the large alliances, the clustering coefficient in the positive and negative networks is comparable and rather stable across the studied time period. In the SOV alliances, on the other hand, the clustering coefficients are larger and more volatile. For the SOV alliances, the positive and negative networks have a sizable different clustering coefficient.  Comparing our EVE Online results for the clustering coefficients with those for  PARDUS~\cite{Szell2010}, we also find a higher clustering coefficient  in the positive networks than in the  negative ones. The reported cluster coefficients in the PARDUS negative subnetwork are of the order of 0.01-0.06, which  is significantly lower than the observed values in EVE Online. We stress that the analysis of the PARDUS network was done on a social network of players, whereas our focus is on a political network of alliances.   

From Fig.~\ref{fig:Network_Prop}, we conclude that the SOV alliances maintain more triadic relations than the +200 ones, that their network is more complete and that they have a stronger tendency to create positive clusters. In the results section we will study how far the +200 and SOV alliances display a more consistent behavior in the way they create the four types of triadic relationships of Fig.~\ref{fig:type_triads}.

\subsubsection{Real World Data: Syrian Civil War and International Relations during the Cold War Era}

\begin{figure*}[!h]
\centering
\includegraphics{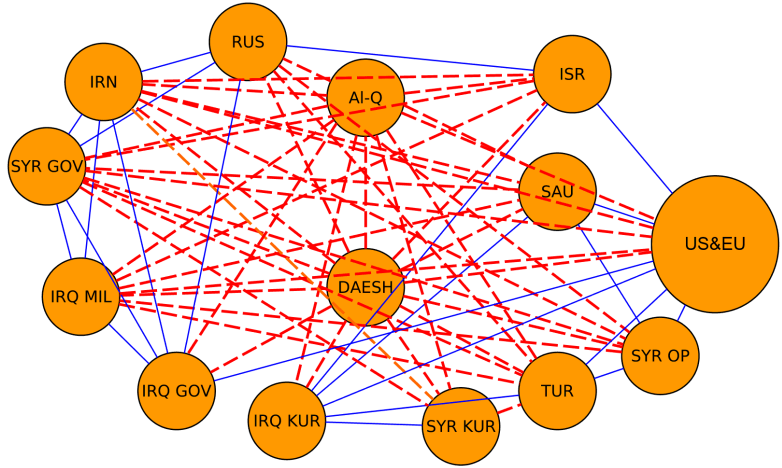}
\caption{\textbf{Network of relationships in the Middle East.} Graph with the relationships between  21 different ``agents'' in the Middle East as reported by The Economist in December 2015 \cite{MiddleEastEconomist}. We discriminate between enemies (red), allies (blue) and neutrals (no link). We use the following abbreviations to specify the nodes: 
Moderate Sunni Arab opposition (SYR OP);
Syrian government (SYR GOV); 
Syrian Kurds (SYR KUR);
Iraqi Kurds (IRQ KUR); 
Iraqi Shia militias (IRQ MIL);
Iraqi government (IRQ GOV);
Turkey (TUR);
Israel (ISR);
Russia (RUS);
USA and E.U (US\&EU); 
Saudi Arabia \& Arab League (SAU);
Iran (IRN);
Al-Qaeda/Jabhat al-Nusra (AL-Q);
DAESH or ISIS (DAESH). The node ``US\&EU'' represents 8 agents (US, Australia, Belgium, Canada, France, Germany, Netherlands and United Kingdom).}
\label{fig:MiddleEast}
\end{figure*}

Here we describe our application of the proposed methodology to the triadic relations of two real-world systems. First, we analyze the triadic relations between 21 \dblquote{agents} in the Middle East in the military intervention against ISIS (or DAESH) as reported by The Economist~\cite{MiddleEastEconomist} on December 23, 2015.  The corresponding graph is displayed in Fig.~\ref{fig:MiddleEast} and is complete in the sense that each possible edge exists with a valence of either ``friendly'', or ``unfriendly/enemy'' or ``neutral/mistrust''. The data provided in ~\cite{MiddleEastEconomist} aggregates all the information about the status of the relationships of the European countries, the US and Australia in one single \dblquote{US \& EU} node.  In order to reach a higher level of detail for the occupation probabilities of the different types of triads, we duplicate the \dblquote{US \& EU} node and introduce nodes (extra agents) for the US, Australia, Belgium, Canada, France, Germany, Netherlands and U.K, as countries that are involved in the conflict \cite{Wiki}.

As a second system, we study the time series of signed edges of international relations during the Cold War era (1949-1989) from the Correlates of War project \cite{COW, AlliancesDatabase,MIDs}. The edge between two countries is considered ``$+$'' whenever there is an active military alliance or defense treaty between them, and we retrieved this data from the dataset for Formal Alliances (version 4.1)~\cite{AlliancesDatabase}.

Correlates of War maintains a database of Militarized Interstate Disputes (MIDs) ~\cite{MIDs}, and we assign a ``$-$'' value to the edges between countries whenever they were involved in an unresolved MID in a time window of 50 years. Countries can resolve their disputes and forge an active military alliance and/or a defense treaty. We mark this as a transition from a ``$-$'' to a ``$+$'' edge between countries. This procedure gives rise to a degeneracy of two for the positive edges and of one for the negative ones.  
As an illustrative example of how we construct the status of international relations we present the case of USA and Germany.  In 1943, both countries are involved in a MID, implying the start of a period of a ``$-$'' relationship. In 1950, an alliance between West Germany and the USA comes into being and their relationship changes from ``$-$'' to ``$+$''.  This does not happen with East Germany which implies that in 1950 its relationship with the USA remains ``$-$''.  

\section{Results}
\subsubsection{Triadic Relations between Alliances in EVE Online}

As pointed out in the previous section and summarized in Table~\ref{table:Levels}, the four types of triadic relationships can be  characterized by a degeneracy and an energy. The lower the energy the more stable the triadic relationship is perceived. In the undirected network formulation of SBT, the degeneracy is purely geometric (Table~\ref{table:Levels}). In EVE Online, we have an extra multiplicity, associated with the conversion rules (Table~\ref{table:PTM}) used to convert the  directed  into undirected edges. The total degeneracy $g$ is the product of the geometrical degeneracy $g_G$ and the multiplicity (see Table~\ref{table:PTM})  associated with each of the 3 edges in a triad.  As can be inferred from Table~\ref{table:PTM} the aggregated weight of the \dblquote{$+$} relationships is 5 and for \dblquote{$-$} it is 7. For example, for the $[+--]$ configuration, the total degeneracy is $g (E_D) = 3  \times \left( 5 \times 7 \times 7 \right)= 735$. 

\begin{figure*}
\centering
\includegraphics[scale=0.8]{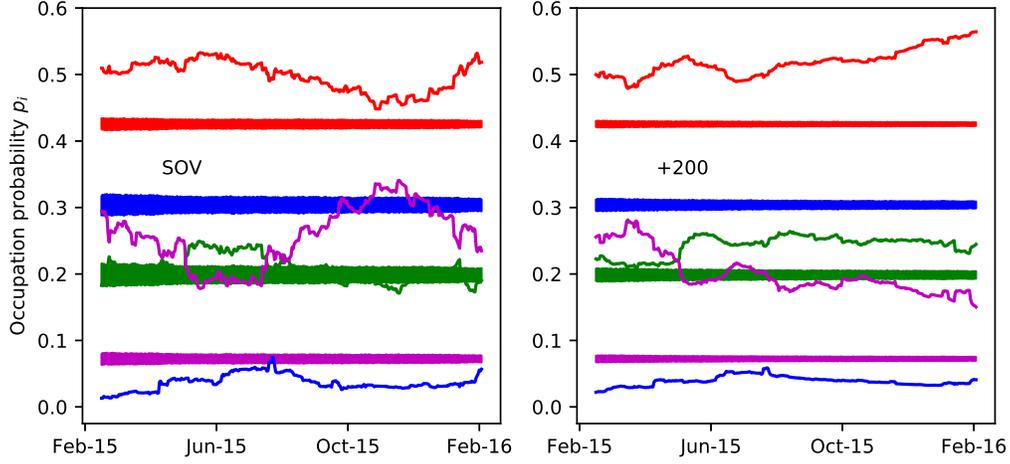}
\caption{\textbf{Comparison between a randomly signed network and EVE's real network of alliances.}
Daily changes of the occupation probabilities for the four types of triads for the relationships between the alliances in EVE Online.  We compare the data with a randomly signed directed network. Blue: $[++-]$; green: $[---]$; red: $[+--]$;  magenta: $[+++]$. The left (right) panel is for alliances with sovereignty  (alliances with more than 200 members).
}
\label{fig:histogram}
\end{figure*}

In Fig.~\ref{fig:histogram}, the time series of the occupation probabilities defined in Eq.~(\ref{eq:occupprobability}) is shown. We compare the data extracted from the alliances in EVE Online with those from a randomization whereby the following procedure has been followed. At each time step in the data we fix the topology of the network as it appears in the data. In a next step, we randomly assign the values $s_{i\rightarrow j} \in \{+,-\}$ and $s_{j \rightarrow i} \in \{+,-\}$ to the edges. Next, we use the rules of Table~\ref{table:PTM} to obtain the status of the edges in the associated undirected network and determine the occupation probabilities $p_i(t)$ for the realization of the network at a particular time step. The procedure of randomly assigning a status to the directed edges is repeated 500 times at each time step in the data. This produces the bands for the $p_i(t)$ in Fig.~\ref{fig:histogram}. 
The comparison between the occupation probabilities of the simulations and the data from the alliances in EVE Online, displays clear patterns that are persistent over time. The stochastic networks produce a higher occupation of the classical unbalanced $[++-]$ and $[---]$ triads than the data. The rate of incidence of the other types of triads is systematically higher in the data than in the simulations. This clearly illustrates that there is dynamics at play and that randomness cannot give rise to the observed patterns in the occupation probabilities. Remarkably, we observe the same ordering of the configurations as seen in the PARDUS data \cite{Szell2010}. In the EVE Online virtual world, Fig.~\ref{fig:histogram} shows that the balanced $[+++]$ triad are on average less populated than the unbalanced  $[---]$ one. This illustrates that the total degeneracies $g$ are a key element in the outcome of the relative occupation probabilities of the different triads.

\begin{figure*}
%
\begin{adjustwidth}{-0.9in}{0in} 
\centering
\includegraphics[scale=0.8]{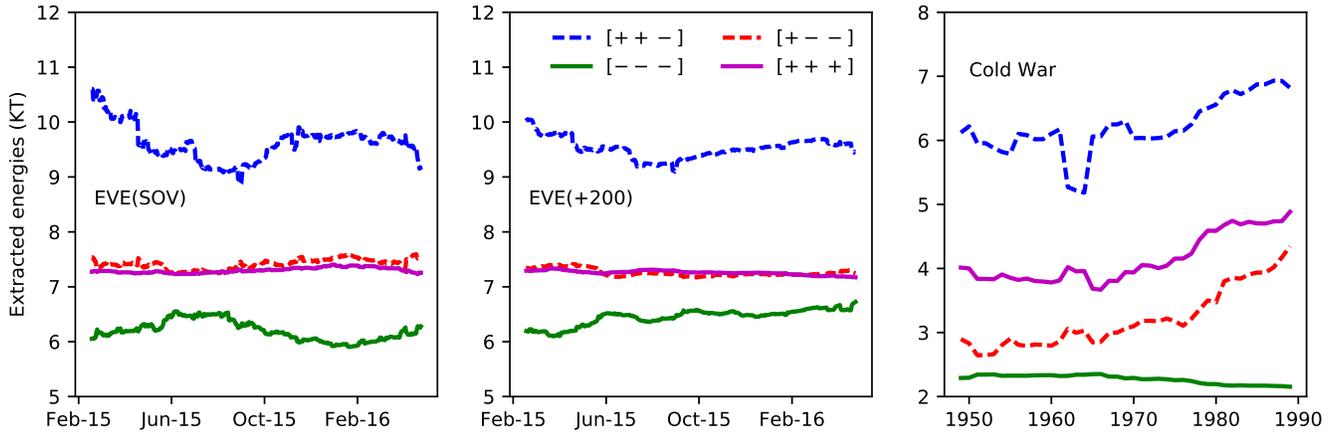}
\end{adjustwidth}
\caption{ \textbf{Time series of the extracted triadic energies.} We display the daily values of the log occupation probabilities $-\ln p_i/g (E_i)$ for the four types of triadic relationships between the +200 (left) and SOV (middle) alliances in EVE Online. These results are based on the data of Fig.~\ref{fig:histogram}. The right panel displays the yearly values of the $-\ln p_i/g (E_i)$ for the four types of triadic relationships in the international relationships during the Cold War era.
}
\label{fig:evolution_energy}

\end{figure*}
The actual values of the energies  $E_i$ are more informative with regard to the underlying dynamics. Eq.~\ref{eq:Energy_from_ocupation_probability} 
connects the occupation probability $p_i$ for each triadic state to its energy $E_i$  and the \dblquote{temperature}.  Fig.~\ref{fig:evolution_energy} shows the time series of the extracted energies (as $\beta E_i + \ln{Z}$). For both classes of alliances the same ordering of the energy levels emerges. This is a remarkable result given the substantial variations and fluctuations in the number of nodes and clustering properties of the political network of alliances as time progresses (Figs~\ref{fig:nodes_evolution} and \ref{fig:Network_Prop}). 

From the time series of Fig.~\ref{fig:evolution_energy} one can infer information about the relative triadic energies. The extracted values for the $\beta E_i + \ln Z$ are summarized in Table~\ref{table:Exponential_EVE}. Both the hierarchy and the values of the  triadic energies $\beta E_i + \ln Z$ are comparable for the +200 and SOV alliances. The highly balanced triad $[+++]$ has the lowest energy. The energy gap to the second balanced state $[++-]$ is somewhat larger for the SOV alliances. The highly frustrated $[++-]$ triad has the high energy, reflecting its perceived outspoken instability.  The balanced $[+--]$  and unbalanced $[---]$ triads have a comparable energy. 

\begin{table*}[tb]
\begin{flushleft}
\caption{\textbf{Extracted triadic energies.} \label{table:Exponential_EVE}}
\begin{tabular}{ l | c c | c c c c }
Type of triad  & \multicolumn{2}{c|}{Total degeneracy} & 
\multicolumn{4}{c}{$-\ln \frac {p_i} {g(E_i)} = \beta E_i + \ln Z$  } \\ 
            & EVE & Cold War & EVE (SOV) & EVE (+200) & Middle-East & Cold War \\ \hline
%
A: $[++-]$   & $g(E_A)=525$ & $g(E_A)=12$ & $ 9.62 \pm 0.33 $ & $ 9.52\pm0.18 $
& 4.47 & $6.38\pm0.49$ \\ 
B: $[---]$   & $g(E_B)=343$ & $g(E_B)=1$ & $ 7.41 \pm 0.09 $ & $ 7.26\pm0.07 $
& 2.11 & $3.24\pm0.39$ \\  
D: $[+--]$ 	 & $g(E_D)=735$ & $g(E_D)=6$ & $ 7.30 \pm 0.04 $ & $ 7.26\pm0.03 $
& 1.80 & $4.25\pm0.40$ \\  
C: $[+++]$   & $g(E_C)=125$ & $g(E_C)=8$ & $ 6.17 \pm 0.18 $ & $ 6.45\pm 0.14 $
& 1.06 & $2.24\pm0.06$ \\ 
\end{tabular}
\end{flushleft}
The triadic energies and corresponding error bars are obtained by time averaging the quantity $-\ln p_i/g(E_i) = \beta E_i + \ln \;Z $ for the  alliances (SOV and +200) in EVE Online (left and middle panel of  Fig~\ref{fig:evolution_energy}) and for the international relations during the Cold War era (right panel of Fig~\ref{fig:evolution_energy}).
\end{table*}

\begin{figure*}
\centering
\includegraphics[scale=0.8]{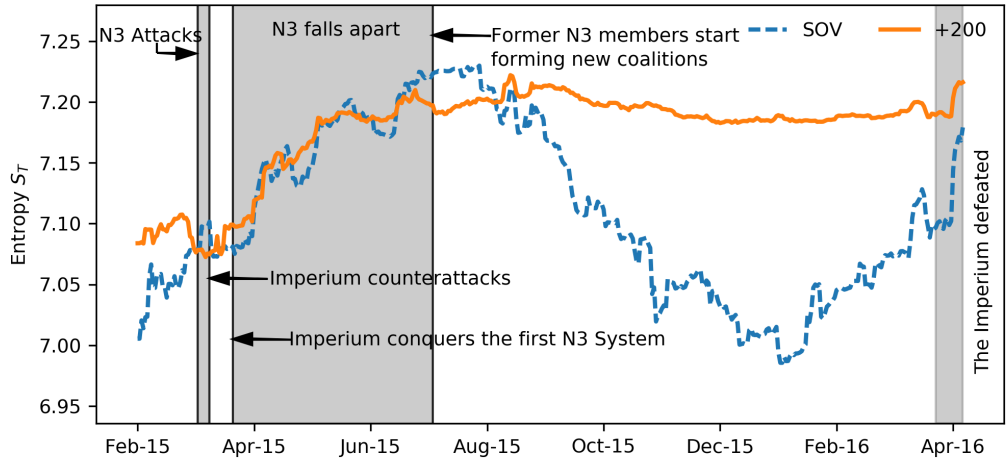}
\caption{\textbf{Time series of the entropy for the triadic relationships in EVE Online.}
Daily values of the entropy  $S_T$ for the occupation probability of the four types of triads in the relationships between the SOV (dashed line) and the +200 (solid line) alliances in EVE Online. The entropies associated with the networks of randomly assigned edges are $S_T= 8.152 \pm 0.011 K$ (SOV) and $S_T= 8.152 \pm 0.066K$ (+200).}
\label{fig:evolution_entropy_EVE}
\end{figure*}

Fig.~\ref{fig:evolution_entropy_EVE} shows the time series of the systemic entropy values $S_T$  defined in Eq.~\ref{eq:entropy_occupation}. In the networks with randomly assigned edges (see Fig.~\ref{fig:histogram}) the time-averaged systemic entropies are $S_T= 8.152 \pm 0.011 K$ (SOV) and $S_T= 8.152 \pm 0.066K$ (+200). 
Obviously, the systemic entropy corresponding with the occupation probabilities for the alliances in EVE Online is substantially smaller. This indicates that even in the worst periods with a lot of turmoil and tensions in EVE Online, the system is still quite a bit more organized than random. 

The time series of the entropy $S_T$  could potentially teach us about the change-points in the time series of triadic relations in the political network. Strong variations in the time series of $S_T$ for the SOV alliances occur in March 2015 and in April 2016. From the documented history of EVE Online, it is known that in 19-24 March 2015, a large coalition of collaborating alliances known as the ``N3'' tried to invade the space controlled by the ``The Imperium''. After this invasion failed, the Imperium counterattacked and N3 fell apart during the following months. The gradual disintegration of N3 is reflected in a continuous increase of entropy (more randomness in the system). This rise of the entropy comes to an end during August-September 2015 by which time the N3 alliances had formed new coalitions. Another distinctive feature in the time series of the entropy is the spike in April 2016. This marks the next great war, known as ``World War Bee'' or ``The Casino War'', that struck EVE's virtual world. As a result of this war, The Imperium coalition fell apart in mid April 2016. This is clearly visible as a strong rise in the $S_T$ for both types of alliances.

\subsubsection{Real World data: Syrian Civil War and International relations during the Cold War}

\begin{figure*}
\centering
\includegraphics[scale=0.9]{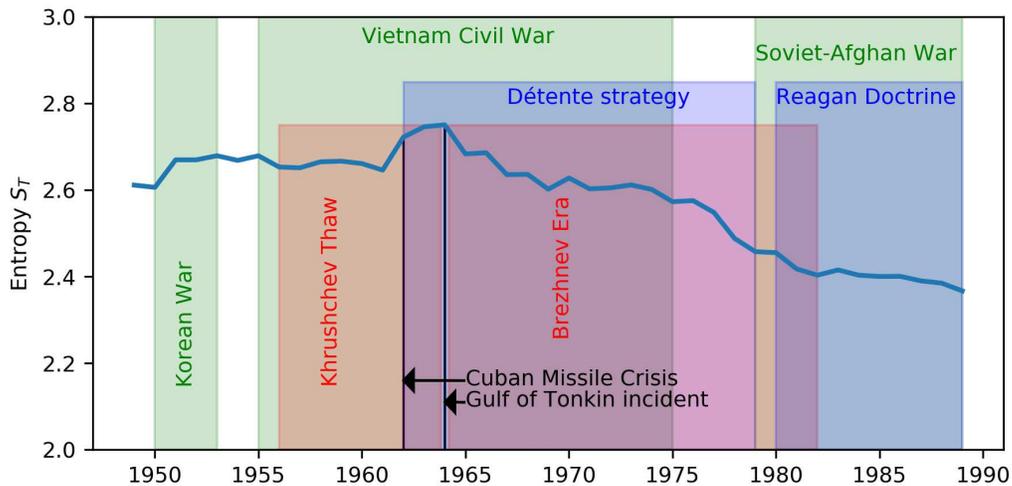}
\caption{\textbf{Time series of the entropy for the triadic relationships among countries during the Cold War era.}}
\label{fig:evolution_entropy_Real_World}
\end{figure*}

The information about the relationships of 21 agents involved in the Syrian Civil of Fig.~\ref{fig:MiddleEast} can be converted into the occupation probabilities $p_i$ for the four types of triads. Using the geometrical degeneracies $g_C(E_i)$ of Table~\ref{table:Levels}, we obtain the $\beta E_i + \beta\ln Z$ values as contained in Table~\ref{table:Exponential_EVE}. Using data from the Correlates of War project \cite{COW} we extracted the times series for the triadic energies (Fig.~\ref{fig:evolution_energy}) and entropies (Fig ~\ref{fig:evolution_entropy_Real_World}). 
The highest entropy value occurred during the 1962-1964 period that marks the Cuban Missile Crisis (October 16–28, 1962) and the Gulf of Tonkin incident (1964) that triggered the USA intervention in Vietnam. The impact of those incidents is also clearly visible in the extracted energies (right panel of Fig.~\ref{fig:evolution_energy}) which show a decreasing trend from 1964 onwards. This trend can be attributed to a systematically increasing occupation probability for the $[+ + +]$ triads at the cost of the occupation probabilities for triads including a "$-$" edge.
This is likely the result of the easing of the strained relations between the blocks of countries (often referred to as the ``d\'etente''). 
Thus one can identify major change points in each political system from the changes in the entropy over time, which the direction of the changes intuitively translating into increased of decreased political stability.
 
\subsection{Model Hamiltonian for SBT }

The extracted triadic energies $\overline{\beta E_i + \ln Z} \pm \sigma_{\beta E_i}$ shown in Table~\ref{table:Exponential_EVE} can be used to determine the strength parameters for the Hamiltonian of Eq.~\ref{eq:hamiltonian}, as well as the unknown zero level $Z_0$. From this we can learn about the underlying dynamics in the formation of the triadic relationships. We use the maximum likelihood method to  determine the $\left(\alpha, \gamma,\omega,Z_0\right)$ that has the highest  probability of generating the extracted $\beta E_i + \ln Z$ sets. The likelihood is computed from 

\begin{align}
\begin{split}
\mathcal{L}(& \beta E_i + \ln Z|\alpha,\gamma, \omega, Z_0) = \prod_{i\in(A,B,D,C)} \\ &
\exp \left(
\frac{-\biggl( \; \overline{\beta E_i+\ln{Z}} -( \beta H_i(\alpha,\gamma,\omega)+Z_0)\biggr)^2}{\sigma^2_{\beta E_i}} 
\right) \; 
\label{eq:likelihood}
\end{split}
\end{align}

The values for the $\left(\alpha, \gamma, \omega \right)$ are listed in Table~\ref{table:Parameter_value}. Remarkably, similar features and values emerge for the four datasets. The three-body and two-body interactions are clearly the driving forces in the creation of the triadic relationships. The fact that the $\omega$ adopts a positive value indicates that we are dealing with political systems with incentives for enmity. The extracted value of $\gamma$ for the Cold War dataset is significantly higher than in the other three systems. This alludes to a stronger homogenization tendency. A possible explanation is the peculiar political situation during the Cold War era, with 3 blocks of countries (Capitalist, Communist and Non-Aligned Movement). We find that the first two groups maintain a high degree of internal homogeneity in their inter-country relationships and their attitudes toward confrontation between the blocks. The origin of this clustering effect can be attributed to the Cold War military strategy of Mutually Assured Destruction. As a consequence, the countries tend to align themselves with a country possessing nuclear weapons.  

\begin{table*}[tb]
\centering
\begin{tabular}{ l | c c c c }
Dataset  & \multicolumn{4}{c}{Strength parameters in units $KT$}  \\
& $\alpha$ & $\gamma$ & $\omega$ & $Z_0$  \\  \hline
EVE (SOV) &  $0.95\pm0.03$ &  $0.38\pm0.02$ &  $0.18\pm0.02$ &  $8.00\pm0.03$ \\             
EVE (+200) & $1.02\pm0.04$ & $0.41\pm0.02$ & $0.14\pm0.02$ & $8.04\pm0.04$ \\  
Middle-East & 1.09 & 0.38 & 0.22 & 2.70 \\  
Cold War & $0.89\pm0.07$ & $0.61\pm0.07$ & $0.14\pm0.07$ & $4.67\pm0.07$ \\ 
\end{tabular} 
\caption{\textbf{Hamiltonian parameters from a theory-data comparison.} \label{table:Parameter_value} The extracted values for the strength parameters for the +200 and SOV alliances in EVE Online, for the status of the relationships in the Middle East  in December 2015 \cite{MiddleEastEconomist}, and for the international relations during the Cold War era.}
\end{table*}

\begin{table*}[tb]
\centering
\begin{tabular}{ l | c c c c }
Dataset  & \multicolumn{4}{c}{$H_i$ in units $KT$ from Hamiltonian of Eq~(\ref{eq:hamiltonian}) } \\ 
& $H_A: [++-]$ & $H_B: [---]$ & $H_D: [+--]$ 	 & $H_C: [+++]$   \\  \hline
EVE (SOV) & $+1.57\pm0.33$ &  $-0.63\pm0.10$ & $-0.75\pm0.05$ &  $-1.85\pm0.18$  \\             
EVE (+200) & $+1.51\pm0.33$ & $-0.75\pm0.06$ & $-0.75\pm0.04$ & $-1.56\pm0.14$ \\  
Middle-East & +1.69 & -0.71 & -0.93 & -1.55 \\  
Cold War & $+1.64\pm0.44$ &  $-1.36\pm0.46$ &  $-0.42\pm0.38$ & $-2.29\pm0.07$ \\ 
\end{tabular} 
\caption{\textbf{Hamiltonian parameters from a theory-data comparison.} \label{table:Parameter_value_2} The corresponding energies $H_i$ (see Table~\ref{table:Levels}) of the Hamiltonian of Eq~(\ref{eq:hamiltonian}) for the +200 and SOV alliances in EVE Online, for the status of the relationships in the Middle East  in December 2015 \cite{MiddleEastEconomist}, and for the international relations during the Cold War era.}
\end{table*}

\section{Conclusion}

We have proposed a methodology to qualitatively study balance theory for triadic relations in political networks. The crux of our framework is the analysis of the occupation probabilities for the four different types of triads. The model uses elements from  Boltzmann-Gibbs statistical physics to assign an energy and a degeneracy to each type of triad and introduce an overall systemic temperature to account for the disorder generating effects. The energies can be investigated using a generic Hamiltonian with three-body, two-body and one-body interactions between the edges in 3-node cycles. The interactions have their own characteristic strength parameters that can be extracted from a model/data comparison.  

We have tested the underlying assumptions of our model with two datasets from the virtual world EVE Online and two datasets from the real world.  We have demonstrated that the proposed model allows one to quantitatively study social balance and gain insight into the mechanisms driving triadic relationships in political networks. For example, we can separate the dynamical mechanisms (regulated by the values of the energies) and stochastic aspects (degeneracy of the different energy levels).  The model/data comparison for our four political networks lead to comparable energy and strength parameters. We furthermore find a persistent hierarchy among the four types of triads. The $[+++]$ triad is consistently the most balanced and the $[++-]$ the most unbalanced triadic relation whereas the $[+--]$ and $[---]$ triads are comparably stable.  The time series of the Shannon entropy corresponding to the occupation probabilities of the different triads allows one to study the activity of the system and to detect the change points in the time series. For EVE Online, the change points are connected with wars or collapses of clusters of alliances. 

The data that we analyzed clearly indicate that there is no clear tendency toward an increased occupation of the balanced triads as time progresses. This is in line with observations for the triadic relations between countries~\cite{Doreian2015} and can be efficiently captured by the introduction of a finite systemic temperature. The determined values for the strength parameters confirm the importance of SBT, namely that three-body forces play a key role in the formation of the network of relationships. We also find, however, strong corrections via the two-body force. This is indicative for a strong inclination to homogenize the status of the relationships in the triads.

\section{Acknowledgments}
The authors are greatly indebted to E\dh vald G\'islason and Andie Nordgren at CCP Games for their help with extracting the data for EVE Online. We are also grateful for many fruitful discussions with Benjamin Vandermarliere.
 


%
%
%

\bibliographystyle{apalike}
\bibliography{biblio.bib} 

\begin{thebibliography}{}

\bibitem[Mid, 2015]{MiddleEastEconomist}
 (2015).
\newblock Friends and foes: rifts in the middle east.
\newblock The Economist.

\bibitem[COW, 2016]{COW}
 (2016).
\newblock The correlates of war.

\bibitem[Abell, 1968]{Abell1968}
Abell, P. (1968).
\newblock Structural balance in dynamic structures.
\newblock {\em Sociology}, 2(3):333--352.

\bibitem[Abell and Ludwig, 2009]{ABELL2009}
Abell, P. and Ludwig, M. (2009).
\newblock {Structural Balance: A Dynamic Perspective}.
\newblock {\em The Journal of Mathematical Sociology}, 33(2):129--155.

\bibitem[Antal et~al., 2005]{Antal2005}
Antal, T., Krapivsky, P.~L., and Redner, S. (2005).
\newblock Dynamics of social balance on networks.
\newblock {\em Phys. Rev. E}, 72:036121.

\bibitem[Antal et~al., 2006]{Antal2006}
Antal, T., Krapivsky, P.~L., and Redner, S. (2006).
\newblock {Social balance on networks: The dynamics of friendship and enmity}.
\newblock {\em Physica D: Nonlinear Phenomena}, 224(1-2):130--136.

\bibitem[Cartwright and Harary, 1956]{cartwright1956sbg}
Cartwright, D. and Harary, F. (1956).
\newblock {Structural balance: a generalization of Heider's theory.}
\newblock {\em Psychological Review}, 63(5):277--93.

\bibitem[CCP, 2016]{EveOnline}
CCP (1997-2016).
\newblock Eve online.

\bibitem[Davis, 1967]{Davis1967}
Davis, J.~A. (1967).
\newblock Clustering and structural balance in graphs.
\newblock {\em Human Relations}, 20:181--187.

\bibitem[Doreian and Mrvar, 2015]{Doreian2015}
Doreian, P. and Mrvar, A. (2015).
\newblock Structural balance and signed international relations.
\newblock {\em Journal of Social Structure}, 16(50).

\bibitem[Estrada and Benzi, 2014]{Estrada2014}
Estrada, E. and Benzi, M. (2014).
\newblock Walk-based measure of balance in signed networks: Detecting lack of
  balance in social networks.
\newblock {\em Phys. Rev. E}, 90:042802.

\bibitem[Facchetti et~al., 2011]{Facchetti2011}
Facchetti, G., Iacono, G., and Altafini, C. (2011).
\newblock Computing global structural balance in large-scale signed social
  networks.
\newblock {\em Proceedings of the National Academy of Sciences},
  108(52):20953--20958.

\bibitem[Galam, 1996]{Galam1996}
Galam, S. (1996).
\newblock Fragmentation versus stability in bimodal coalitions.
\newblock {\em Physica A: Statistical Mechanics and its Applications},
  230(1):174 -- 188.

\bibitem[Gawronski et~al., 2005]{Gawroski2005}
Gawronski, P., Gronek, P., and Kulakowski, K. (2005).
\newblock {The heider balance and social distance}.
\newblock {\em Acta Physica Polonica B}, 36(8):2549--2558.

\bibitem[Gibler, ]{AlliancesDatabase}
Gibler, D.~M.
\newblock “international military alliances, 1648-2008.”.
\newblock {\em CQ Press}.

\bibitem[Hart, 1974]{Hart1974}
Hart, J. (1974).
\newblock Symmetry and polarization in the european international system,
  1870-1879: a methodological study.
\newblock {\em Journal of Peace Research}, 11(3):229--244.

\bibitem[Heider, 1946]{heiderattitudes}
Heider, F. (1946).
\newblock Attitudes and cognitive organization.
\newblock {\em Journal of Psychology}, 21:107--112.

\bibitem[Hummon and Doreian, 2003]{Hummon2003}
Hummon, N.~P. and Doreian, P. (2003).
\newblock Some dynamics of social balance processes: Bringing heider back into
  balance theory.
\newblock {\em Social Networks}, 25(1):17--49.

\bibitem[Lerner, 2016]{Lerner2016}
Lerner, J. (2016).
\newblock Structural balance in signed networks : separating the probability to
  interact from the tendency to fight.
\newblock {\em Social Networks}, 45:66--77.

\bibitem[Leskovec et~al., 2010]{Leskovec2010}
Leskovec, J., Huttenlocher, D., and Kleinberg, J. (2010).
\newblock Signed networks and in social and media.
\newblock In {\em CHI 2010: Machine Learning and Web Interactions April
  10–15, 2010, Atlanta, GA, USA}.

\bibitem[Marvel et~al., 2009]{Marvel2009}
Marvel, S.~A., Strogatz, S.~H., and Kleinberg, J.~M. (2009).
\newblock Energy landscape of social balance.
\newblock {\em Phys. Rev. Lett.}, 103:198701.

\bibitem[Palmer et~al., ]{MIDs}
Palmer, G., D’Orazio, V., Kenwick, M., and Lane, M.
\newblock “the mid4 data set: Procedures, coding rules, and description.”.
\newblock {\em Conflict Management and Peace Science}.

\bibitem[Press\'e et~al., 2013]{Presse2013}
Press\'e, S., Ghosh, K., Lee, J., and Dill, K.~A. (2013).
\newblock Principles of maximum entropy and maximum caliber in statistical
  physics.
\newblock {\em Rev. Mod. Phys.}, 85:1115--1141.

\bibitem[Saiz et~al., 2017]{Saiz2017}
Saiz, H., Gómez-Gardeñes, J., Nuche, P., Girón, A., Pueyo, Y., and Alados,
  C.~L. (2017).
\newblock Evidence of structural balance in spatial ecological networks.
\newblock {\em Ecography}, 40(6):733--741.

\bibitem[Szell et~al., 2010]{Szell2010}
Szell, M., Lambiotte, R., and Thurner, S. (2010).
\newblock Multirelational organization of large-scale social networks in an
  online world.
\newblock {\em Proceedings of the National Academy of Sciences},
  107(31):13636--13641.

\bibitem[Traag et~al., 2013]{Traag2013}
Traag, V.~A., Van~Dooren, P., and De~Leenheer, P. (2013).
\newblock Dynamical models explaining social balance and evolution of
  cooperation.
\newblock {\em PLOS ONE}, 8(4):1--7.

\bibitem[Vinogradova and Galam, 2014]{Vinogradova2014}
Vinogradova, G. and Galam, S. (2014).
\newblock Global alliances effect in coalition forming.
\newblock {\em The European Physical Journal B}, 87(11):266.

\bibitem["Wikipedia", 2016]{Wiki}
"Wikipedia" (2016).
\newblock Military intervention against isil --- {W}ikipedia{,} the free
  encyclopedia".
\newblock [Online; accessed 22-Apr-2016].

\end{thebibliography}
\end{document}